\documentstyle [12pt] {article}

\begin{document}
\begin{flushright}
IUHET-276\\
April 1994\\
Revised May 1994\\
\end{flushright}
\vspace*{0.5 cm}
\begin{center}
{\bf NEUTRINO FLAVOR EVOLUTION\\
NEAR A SUPERNOVA'S CORE} \\
\vspace*{0.5 cm}
\vspace*{0.5 cm}
{\bf J. Pantaleone}\\
\vspace*{0.5 cm}
Physics Department\\
Indiana University, Bloomington, IN 47405 \\
\vspace*{0.5 cm}
\vspace*{0.5 cm}
{\bf ABSTRACT} \\
\end{center}

In supernovae and in the early universe,
neutrino flavor evolution is a many-body phenomena.
Here the equations describing the evolution of
the density matrices in phase space are derived.
Then these equations are applied to
neutrino emission from a supernova core.
The effects of a `small' background neutrino density
on adiabatic and nonadiabatic flavor evolution are calculated
analytically.
It is found that when flavor evolution is sizeable,
the sensitivity to the small neutrino background is enhanced.
This implies that r-process nucleosynthesis in
supernovae may not reliably probe
neutrino masses less than about 25 eV.

\newpage

Forward scattering can have an enormous impact on neutrino
flavor evolution.  Neutrino forward scattering off of an electron
background is responsible for the Mikheyev-Smirnov-Wolfenstein
(MSW) effect \cite{Wolfenstein,MS} which may explain the solar
neutrino problem (for a review, see e.g. \cite{KPr}).
Neutrino forward scattering off of neutrinos \cite{FMWS,NR} \cite{JP}
has been shown to produce curious, coherent phenomena in the early
universe \cite{KPS}.
Here I consider flavor evolution
when the electron and neutrino densities are comparable.
This is the case just outside the core of a supernova
where nucleosynthesis of heavy elements is believed to occur.
Since nucleosynthesis is very sensitive to the relative
neutrino spectra, it may be possible to probe a
range of neutrino masses and mixings \cite{PRL} which
are of great interest as candidates for nonbaryonic dark matter.

This paper is divided into two parts.  First
the general equations describing neutrino flavor evolution
in phase space are {\it derived} in the density matrix
approximation (see also \cite{Dolgov,Stodolosky,KPS,RS,Thomson}).
All of the conceptual steps are given here,
but the mathematical details will be published elsewhere.
These evolution equations have many obvious applications
in the early universe and in supernovae.
In particular, in the second half of the paper they are
solved to find the corrections to level crossing
from a `small' neutrino background.

\section{The evolution equations.}

One starts be making the physically reasonable assumption
that each neutrino is described by its own wavefunction
(the Hartree approximation).
That is, each neutrino obeys a Dirac (or Majorana) equation with
a potential.  The potential for a particular neutrino is
calculated by adding up all possible two-body, forward
scattering, weak interaction amplitudes
between that neutrino and all the other particles present.
Note that these initial two steps
neglect Pauli blocking and so inherently assume
that the neutrinos are dilute.

Then each individual neutrino wave function is factored
into a `quickly' varying part
(the solution to the massless wave equation)
and a `slowly' varying part.
The `quickly' varying part describes the propagation of a
wave packet, and the `slowly' varying part describes how
the wave packet gradually separates into different
wavepackets---one for each mass eigenstate.
Because practical position `measurements' of neutrinos
are always much much coarser than the size of the wave packet
(which is the inverse of the neutrino momentum spread),
these individual equations should be
spatially averaged over a scale intermediate
between the `quick' and `slow' scales.
This eliminates the
interference effects from the `quickly' varying part of the wavefunction,
so that the 3-dimensional
position and momentum vectors are then classical variables.
There remains approximate equations for the `slowly' varying
part of the individual neutrino wave functions.  These describe
the interference effects resulting from the different
neutrino vacuum masses.

The potentials depend on bilinears of the fermion wavefunctions,
i.e. particle densities.  With a little algebra, the equations for
the individual neutrino wavefunctions are easily
converted to equations for the bilinears
\begin{eqnarray}
\rho_i(\vec{x},t,\vec{k}) & = &
\left[ \begin{array}{c} \nu_e \\ \nu_\mu \\ \nu_\tau \end{array} \right]
 \left[ \begin{array}{c c c} \nu_e^\ast & \nu_\mu^\ast &
 \nu_\tau\ast \end{array} \right]_i(\vec{x},t,\vec{k})  \nonumber \\
\bar{\rho}_i(\vec{x},t,\vec{k}) & = &
 \left[ \begin{array}{c} \bar{\nu}_e \\ \bar{\nu}_\mu \\
  \bar{\nu}_\tau \end{array} \right]
  \left[ \begin{array}{c c c} \bar{\nu}_e^\ast & \bar{\nu}_\mu^\ast &
  \bar{\nu}_\tau^\ast \end{array} \right]_i(\vec{x},t,\vec{k})
\end{eqnarray}
Here \( \nu_\alpha \) is the slowly varying part of the
ith neutrino's wavefunction
of flavor \(\alpha\) with momentum \(\vec{k}\) at time \(t\) and
position \(\vec{x}\)  (a bar over a quantity always denotes that
it refers to an antineutrino).
The individual evolution equations for a given \(\vec{x}, t\) and \(\vec{k}\)
are identical.
Thus I can sum the \(\rho_i\)'s and \(\bar{\rho}_i\)'s
over all neutrinos with momentum
\(\vec{k}\) at time t in the `small' volume at position \(\vec{x}\)
(over which the full equations where averaged)
to get the density matrices
\begin{equation}
\rho (\vec{x},t,\vec{k}) \equiv \sum \rho_i (\vec{x},t,\vec{k}) \  \ , \  \
\bar{\rho} (\vec{x},t,\vec{k}) \equiv \sum \bar{\rho}_i (\vec{x},t,\vec{k})
\label{sum}
\end{equation}
The integral of \(\rho_{\alpha \alpha} \) over all phase space is
just the total number of neutrinos with flavor \(\alpha\).
The coupled evolution equations for the density matrices are
\begin{equation}
i \left\{ \frac{ \partial }{ \partial t} + \hat{k} \cdot
\frac{ \partial }{ \partial \vec{x} } \right\}
\rho{ ( \vec{x} , t , \vec{k} ) }  =
[ V( \vec{x} , t , \vec{k} ; \rho , {\bar \rho} ) , \rho(\vec{x},t,\vec{k}) ]
\label{neq}
\end{equation}
and
\begin{equation}
i \left\{ \frac{ \partial }{ \partial t} + \hat{k} \cdot
\frac{ \partial }{ \partial \vec{x} } \right\}
\bar{\rho}{ ( \vec{x} , t , \vec{k} ) }  =
[ \bar{V}( \vec{x} , t , \vec{k} ; \rho , {\bar \rho} ) ,
\bar{\rho}(\vec{x},t,\vec{k}) ]
\label{nbeq}
\end{equation}
Here the right hand sides are the commutator of the matrices.
The effective potential matrices \(V\) and \(\bar{V}\) are given
in the flavor basis by
\begin{eqnarray}
\lefteqn{ V(\vec{x},t,\vec{k} ; \rho , {\bar \rho} )  = }   \\
& & \biggl\{ \frac{1}{ 2 k }
U \left[ \begin{array}{c c c} m_1^2 & 0 & 0 \\
0 & m_2^2 & 0 \\ 0 & 0 & m_3^2 \end{array} \right] U^{\dagger} +
\sqrt{2} G_F [ N_e (\vec{x},t) - N_{\bar{e}} (\vec{x},t) ]
\left[ \begin{array}{c c c} 1 & 0 & 0 \\
0 & 0 & 0 \\ 0 & 0 & 0 \end{array} \right]
 \nonumber \\
& &
+ \sqrt{2} G_F \int \frac{ d^3 \vec{q}}{(2 \pi)^3}
(1 - \hat{k} \cdot \hat{q})
\left(
- \bar{\rho}(\vec{x},t,\vec{q})^\ast
+ \rho(\vec{x},t,\vec{q})
\right)
\biggr\} \nonumber
\end{eqnarray}
and
\begin{eqnarray}
\lefteqn{ \bar{V}(\vec{x},t,\vec{k} ; \rho , {\bar \rho} )  = }   \\
& & \biggl\{ \frac{1}{ 2 k }
U^{\ast} \left[ \begin{array}{c c c} m_1^2 & 0 & 0 \\
0 & m_2^2 & 0 \\ 0 & 0 & m_3^2 \end{array} \right] U^{\rm T} +
\sqrt{2} G_F [- N_e (\vec{x},t) + N_{\bar{e}} (\vec{x},t) ]
\left[ \begin{array}{c c c} 1 & 0 & 0 \\
0 & 0 & 0 \\ 0 & 0 & 0 \end{array} \right]
 \nonumber \\
& &
+ \sqrt{2} G_F \int \frac{ d^3 \vec{q}}{(2 \pi)^3}
(1 - \hat{k} \cdot \hat{q})
\left(
\bar{\rho}(\vec{x},t,\vec{q})
- \rho(\vec{x},t,\vec{q})^\ast
\right)
\biggr\} \nonumber
\end{eqnarray}
The first term in the potentials describes the neutrino mixing from
vacuum parameters, where
the \(m_i^2\) are the neutrino vacuum masses and \(U\) is the
leptonic analogue of the Cabibbo-Kobayashi-Maskawa mixing matrix
(see Ref. \cite{KPr} and references therein
for a discussion of issues relevant for 3 neutrino flavors).
The second term in the potentials comes from neutrino
forward scattering off of nonrelativistic
electrons via the weak charged current,
where \(N_e\) (\(N_{\bar{e}}\)) is the electron (positron) number density
\cite{Wolfenstein}.
Neutral current forward scattering of neutrinos
off of electrons, protons and neutrons in
the background gives a contribution proportional to the identity
matrix---which has no effect
on the flavor evolution and so is not shown.

The final term in the V's comes from neutrinos forward scattering
off of background neutrinos and antineutrinos via the weak neutral current
\cite{JP}.
This term does contribute to the flavor evolution
because the 'propagating` and `background' neutrinos and antineutrinos
can exchange flavor.
This term, like all tree level neutral current interactions,
is invariant under
under global flavor transformations of the form
\begin{equation}
\rho ' = W^\dagger \rho W \ \ , \ \
{\bar \rho} ' = W^T {\bar \rho} W^\ast
\end{equation}
where \(W\) is a unitary matrix.
If \(\rho\) and \({\bar \rho}\) are in the flavor basis,
and \(\rho '\) and \({\bar \rho} '\) are in the vacuum mass basis,
then \(W\) is just \(U\).
Note that the \((1 - \hat{k} \cdot \hat{q})\) factor in the
neutrino-neutrino forward scattering term
means that an individual neutrino can not forward scatter off of itself.

These equations describe the effects on neutrino flavor
evolution from background neutrino and electron
densities. They are necessary when neutrino densities are sizeable
such as in supernovae and in the early universe.
However note that nonforward neutrino scattering
has been neglected in these evolution
equations (see \cite{Dolgov,Stodolosky,RS} and references therein).
Thus these equations are appropriate for supernova neutrinos outside
of the neutrinosphere, and for cosmological neutrinos after
neutrino decoupling (where additional CP even terms of order
\(G_F^2\) are sometimes relevant, see \cite{NR}).
Note that these evolution equations
do {\it not} make the two flavor approximation, and
do {\it not} assume that phase space is isotropic.

Under CP symmetry,
\(\rho_\alpha (\vec{x},t,\vec{k}) \leftrightarrow
\bar{\rho}_\alpha (-\vec{x},t,-\vec{k}) \).  The wave equations are
CP symmetric if \(U = U^\ast\) and
\( ( N_e - N_{\bar e} ) = 0\).

The left hand side of Eqs. (\ref{neq}-\ref{nbeq}) is just the total time
derivative of the density matrix (since \(\hat{k}\) is
just the velocity of the wave packet).
When all the quantum interference effects average out,
the right hand side of these equations vanish and
so they reduce to the continuity equations
in phase space, i.e. the Vlasov equation.
The Vlasov equation generally also contains
derivatives of the neutrino density with respect to
momentum.  Such derivatives are also generally present in
these  equations too, in principle.  However changes in neutrino momentum
come from nonforward scattering, or refraction from density
variations, or from gravitational redshift.
These effects are higher order and have been consistently neglected above.

The right hand side of Eqs. (\ref{neq}-\ref{nbeq}) vanishes when one
takes the trace of the matrix.
The traces of \(\rho (\vec{x},t,\vec{k}) \) and
\({\bar \rho} (\vec{x},t,\vec{k}) \) are
the flavor independent number densities in phase space and these
are conserved quantities---they satisfies the continuity
equation in phase space.
The trace of V also does not contribute
to the right hand side of these equations.
Thus in studying neutrino flavor evolution, I can always choose
to work with a traceless \(\rho\) and \({\bar \rho}\).
In fact, it is easily shown that Tr[\(\rho^2 (\vec{x},t,\vec{k}) \)] and
Tr[\({\bar \rho}^2 (\vec{x},t,\vec{k}) \)] are also conserved quantities.
This is because
dissipative effects such as nonforward scattering have not been included
in the evolution equations, so entropy is conserved \cite{Thomson,RS}.
Taking into account all of the constraints,
the flavor content of the three-flavor density matrices
depends on seven independent, real variables at each point
in phase space.
In the two-flavor approximation the density matrices depend on only
two independent, real variables.

It is usually true that the neutrinos are created at high densities,
where the vacuum neutrino masses are negligible compared to the
induced potentials.  Then the neutrinos are approximately in flavor states.
Because the neutrinos are generally produced over times and positions
which are distributed much more broadly than the scale set by
the inverse of the potential, the flavor off diagonal terms generally
average out in the summation of Eq. (\ref{sum}).
A density matrix diagonal in flavor space is a solution to
Eqs. (\ref{neq}-\ref{nbeq}) when the neutrino masses are negligible.
This is typically the appropriate initial condition for neutrinos created
at high densities.

With the simplifying approximation that there are only two
neutrino flavors, the effective potentials in the flavor basis
take the form:
\begin{eqnarray}
\lefteqn{ V(\vec{x},t,\vec{k} ; \rho , {\bar \rho} )  = }
\label{V2} \\
& & \biggl\{ \frac{\Delta}{ 4 k }
\left[ \begin{array}{c c } -C_{2\theta} &  S_{2\theta} \\
S_{2\theta} & C_{2\theta} \end{array} \right] +
\frac{\sqrt{2}}{2} G_F [ N_e (\vec{x},t) - N_{\bar{e}} (\vec{x},t) ]
\left[ \begin{array}{c c} 1 & 0 \\
0 & -1  \end{array} \right]
 \nonumber \\
& &
+ \sqrt{2} G_F \int \frac{ d^3 \vec{q}}{(2 \pi)^3}
(1 - \hat{k} \cdot \hat{q})
\left(
- \bar{\rho}(\vec{x},t,\vec{q})^\ast
+ \rho(\vec{x},t,\vec{q})
\right)
\biggr\} \nonumber
\end{eqnarray}
and
\begin{eqnarray}
\lefteqn{ \bar{V}(\vec{x},t,\vec{k} ; \rho , {\bar \rho} )  = }
\label{Vb2} \\
& & \biggl\{ \frac{\Delta}{ 4 k }
\left[ \begin{array}{c c } -C_{2\theta} &  S_{2\theta} \\
S_{2\theta} & C_{2\theta} \end{array} \right] +
\frac{\sqrt{2}}{2} G_F [- N_e (\vec{x},t) + N_{\bar{e}} (\vec{x},t) ]
\left[ \begin{array}{c c} 1 & 0 \\
0 & -1  \end{array} \right]
 \nonumber \\
& &
+ \sqrt{2} G_F \int \frac{ d^3 \vec{q}}{(2 \pi)^3}
(1 - \hat{k} \cdot \hat{q})
\left(
\bar{\rho}(\vec{x},t,\vec{q})
- \rho(\vec{x},t,\vec{q})^\ast
\right)
\biggr\} \nonumber
\end{eqnarray}
here
\(C_{2\theta} = \cos 2 \theta\),
\(S_{2\theta} = \sin 2 \theta\), where \(\theta\) is the vacuum mixing
angle, and \(\Delta = m_2^2-m_1^2\).

\section{Supernovae Nucleosynthesis.}

In a recent Letter \cite{PRL}, Qian {\it et al.}
showed how the nucleosynthesis of heavy elements
in type II supernovae is sensitive to neutrino masses and mixings.
Nucleosynthesis would not occur if
the \(\nu_e\) and the \(\nu_\mu\) or \(\nu_\tau\) fluxes
emitted from the supernova core were interchanged
before reaching the nucleosynthesis region.
Such a flavor interchange occurs when a massive neutrino
propagates adiabatically through an MSW resonance.
Qian {\it et al.} calculated that this occurs for
neutrino mass-squared differences in the range of
4 eV\(^2\) to \(10^4\) eV\(^2\) when
neutrino mixings are of order
\(10^{-5} \leq \sin^2 2 \theta\).
This result is extremely interesting since it covers
the range of neutrino masses that give a significant
contribution to the cosmological energy density.
Here the effects of the small background neutrino density
on the flavor evolution are studied using Eqs. (\ref{V2}-\ref{Vb2}).

It is reasonable to take the supernova to be spherically symmetric.
Then the density matrix can be parametrized in terms of only 4 quantities:
the magnitude of the neutrino momentum, k, the radial distance, r,
the angle between the neutrino velocity and the radial
direction, \(\phi\), and t.
In addition, the time dependence in the electron background
is negligible since the bulk transport of normal
matter is nonrelativistic.  Thus it is probably reasonable to neglect
the time dependence everywhere and look for static solutions; i.e.
\(\rho = \rho(r,k,\phi)\) and
\({\bar \rho} = {\bar \rho}(r,k,\phi)\).

At first glance, numerical simulation would appear to
be a straightforward way to study the flavor evolution.
However as discussed above there are 3 parameters that must be
simultaneously accounted for.
In addition, one is interested here in
what occurs for very small mixing angles---which requires very
small grid spacings for these three parameters in order
to resolve the MSW resonances.
Both of these considerations make a numerical analysis
extremely more challenging than anything previously
attempted  (see e.g. \cite{KPS}).
Here a different approach is adopted.

I start by assuming that the contribution of the neutrino density
to the potential is negligible to leading order.
Then the leading order evolution equations are the usual,
linear equations that have been well studied
(see e.g. \cite{KPr}).  I then solve these equations
and substitute this leading order solution into the full potentials,
Eqs. (\ref{V2}-\ref{Vb2}),
to get a potential correct to first order in the neutrino density.
With this corrected potential,
the evolution equations are again linear and may be easily solved to
give the first order effects due to a small neutrino background.

At leading order, expressions for the explicit density matrix
solutions can be written down.
The antineutrinos have little flavor evolution
since their energy levels are well separated.
Assuming the antineutrinos are produced at high densities,
the general, traceless solution for the antineutrino density matrix in the
flavor basis is given by
\begin{eqnarray}
{\bar \rho}_0 (r,k,\phi) = {1 \over 2} [ f_{\bar e} - f_{\bar \mu} ]
\left[ \begin{array}{c c}  \cos 2 {\bar \theta}_m &
- \sin 2 {\bar \theta}_m  \\
- \sin 2 {\bar \theta}_m  & - \cos 2 {\bar \theta}_m   \end{array} \right]
\label{rb}
\end{eqnarray}
Here the \(f_\alpha\)'s are just the phase space densities
as calculated for neutrinos that do not mix.  For example,
for a blackbody spectrum
\begin{equation}
f_\alpha (k) = {1 \over [ \exp{(k/T_\alpha)} + 1 ] }
\end{equation}
where \(T_\alpha\) is the temperature of neutrino species \(\nu_\alpha\).
Also, \(\theta_m\) is the leading order neutrino mixing angle in matter
\begin{equation}
\tan 2 \theta_m (k,r) =
{\Delta \sin 2 \theta \over
[- 2 \sqrt{2} G_F N_e (r) k + \Delta \cos 2 \theta ] }
\end{equation}
and for antineutrinos \({\bar \theta}_m\) is given by the same expression
with \(N_e \rightarrow -N_e\).
There are no explicit phases in Eq. (\ref{rb}) because they
average out in the summation of Eq. (\ref{sum}) due to the
variation in initial production positions.
However neutrinos go through a resonance and
so have much larger, and more complicated, flavor evolution.
In the flavor basis the diagonal element of the traceless neutrino density
matrix is
\begin{eqnarray}
\rho_0 (r,k,\phi )_{e e} &=&
 {1 \over 2} [ f_e - f_\mu ]
\{ -  \cos 2 \theta_m [1 - 2 P_c ]  \\
& & \mbox{} + 2  \sqrt{P_c (1-P_c)} \sin 2 \theta_m
\cos(\alpha + \beta) \} \nonumber
\end{eqnarray}
and the off diagonal element is
\begin{eqnarray}
\rho_0 (r,k,\phi )_{e \mu} &=&
 {1 \over 2} [ f_e - f_\mu ]
\{   \sin 2 \theta_m [1 - 2 P_c ] \\
& & \mbox + 2 \sqrt{P_c (1-P_c)}
[ \cos 2 \theta_m \cos(\alpha + \beta) + i \sin(\alpha+\beta) ] \}
\nonumber
\label{ro}
\end{eqnarray}
For \(\sin^2 2 \theta << 1\),
the asymptotic\footnote{The LSZ expression
must be amputated above the resonance \cite{KPr}
but it can reasonably be used below the resonance.}
crossing probability is
well described by the Landau-Stueckelberg-Zener \cite{LSZ}
crossing probability
\begin{equation}
 P_c ( k,\phi) \simeq
\exp \left[ - \gamma   {\pi \over 2}  \right]
\label{LSZ}
\end{equation}
where
\begin{equation}
\gamma \equiv { \Delta \sin^2 2 \theta \over
2 E \cos 2 \theta | dN_e / N_e ds |_0 }
\label{gamma}
\end{equation}
is the adiabaticity parameter
and \(s = \hat{k} \cdot \vec{x}\) is the distance along
the neutrino path.   Also
\(\alpha \equiv (s - s_0) \Delta/2k\) is the phase
generated by vacuum oscillations {\it as measured from the resonance point},
and \(\beta\) is a `topological' phase
acquired by going through the resonance \cite{PP}.

The phases in the leading order neutrino density matrix
make it difficult to calculate the first order corrections.
However the phases vanish and evaluation becomes simple
in the adiabatic, \(P_c = 0\), and nonadiabatic, \(P_c = 1\), limits.

\subsection{The Nonadiabatic limit.}

When the neutrino vacuum mixing is extremely small
the flavor evolution is extremely small.
Then the leading order density matrices are
just diagonal in the flavor basis, as is the \(\nu - e\) potential.
The effect of the neutrino background can then always be incorporated by
defining an effective electron density.
Performing the integration over background neutrinos gives
the effective neutrino density \(N_{NA}\).
Assuming a common neutrinosphere and ignoring limb darkening
\begin{eqnarray}
N_{NA} (r,\phi) & = &
F(\phi,r)
\left[ N_{\nu_e}  - N_{\nu_\mu} -  N_{{\bar \nu}_e} + N_{{\bar \nu}_\mu}
\right]
\end{eqnarray}
Here \(N_{\nu_\alpha}\) is the total number density
of \(\nu_\alpha\) at the neutrinosphere,
e.g. for a blackbody \(N_{\nu_\alpha} = 0.0913 T_\alpha^3 \).
The geometric factor
\begin{equation}
F(\phi,r) = {1 \over 4} [ 2 - (1+x) \cos \phi ] [ 1 - x ]
\end{equation}
describes the
change in the effective neutrino flux outside of the neutrinosphere.
Here \(x \equiv \sqrt{1 - (R/r)^2}\),
and R is the radius of the neutrinosphere
(see \cite{GDN} for details on similar geometric
factors).\footnote{When \(F\) is averaged over \(\phi\),
and taking the limit of \(r >> R\), then
\( < F >  \approx {1 \over 8 } \left[ { R / r } \right]^4 \).}

In the nonadiabatic limit,
\( P(\nu_e \rightarrow \nu_\mu ; k,\phi) \simeq \gamma {\pi \over 2 }
\) for \(\sin^2 2 \theta << \gamma << 1\) and
a small neutrino background is fully accounted for by using this expression
with the replacement
\(N_e ( r ) \rightarrow N_e ( r ) + N_{NA} ( r, \phi) \)
in \(\gamma\), Eq. (\ref{gamma}).
In supernovae the \(N_{NA}\)
has the same sign as the electron density \(N_e\)
and is at most 20\%-30\% of its size \cite{PRL}.
The rate of change of \(N_{NA}\) only slightly exceeds
that of \(N_e\) at distances far from the core.
Thus the effects of the neutrino-antineutrino background on neutrino
flavor evolution near the nonadiabatic limit
are generally relatively minor.
This is apparently the general situation the
authors of Ref. \cite{PRL} envisioned.

\subsection{The Adiabatic limit.}

However, for almost all of the parameter region probed by supernovae
nucleosynthesis, the flavor evolution is
extremely adiabatic at leading order
in the small neutrino background.
Substituting the leading order density matrices
with \(P_c = 0\) into the potentials
and performing the integration over background momentum,
the diagonal elements of the potential in the flavor basis are
\begin{eqnarray}
[V_{\mu \mu}
 - V_{e e}](r, k,\phi; \rho_0, {\bar \rho}_0 )  =
{ \Delta \over 2 k } \cos 2 \theta - \sqrt{2} G_F
[ N_e(r) +  N_{Ad} (r,\phi) ]
\label{Vmm}
\end{eqnarray}
where, to leading order in \(\sin 2 \theta\),
\begin{eqnarray}
N_{Ad} (r,\phi) & \simeq &
F(\phi,r)
\left[
-N_{\nu_e} (q<q_r) + N_{\nu_e} (q>q_r) + N_{\nu_\mu} (q<q_r)
\right. \nonumber \\
& & \mbox{} \ \ \  \ \ \ \
\left. - N_{\nu_\mu} (q>q_r) + N_{{\bar \nu}_\mu} - N_{{\bar \nu}_e}
\right]
\end{eqnarray}
Here \(N_{\nu_\alpha}(q<q_r) \) (\(N_{\nu_\alpha}(q>q_r) \)) is the
number density of \(\nu_\alpha\) in equilibrium at the neutrinosphere
that have momenta less than (greater than) the resonance momentum
at radius r, \(q_r \equiv \Delta \cos 2 \theta
/ [2 \sqrt{2} G_F N_e(r) ]\).
The `reason' for the inequalities in \(N_{Ad}\) is straightforward,
the flavor content of the background neutrinos depends on whether or not
they have gone through their resonances.
Similarly, the off-diagonal element in the flavor basis is
\begin{eqnarray}
{V_{e \mu}(r, k,\phi; \rho_0, {\bar \rho}_0 ) = }
{1\over 2} \sin 2 \theta \ \ [ {\Delta \over 2 k}  -
\sqrt{2} G_F N_{Ao}(r,\phi) ]
\label{Vem}
\end{eqnarray}
where, to leading order in \(\sin 2 \theta\),
\begin{eqnarray}
N_{Ao} (r,\phi) & \simeq &
F(\phi,r)
\left[
2 \ln | 2 \cot 2 \theta |  q_r
( {dN_\mu \over d q}|_{q_r} -  {dN_e \over d q}|_{q_r} )
\right. \\
& & \mbox{}
\left.
-N_{\nu_e} (q<q_r) + N_{\nu_\mu} (q<q_r)
+ N_{{\bar \nu}_\mu}(q<q_r) - N_{{\bar \nu}_e} (q<q_r)
\right] \nonumber
\end{eqnarray}
Here \(dN_\alpha / dq|_{q_r} \equiv q_r^2 f_\alpha (q_r)/(2 \pi^2) \)
is the derivative of the
number density of \(\nu_\alpha\) in equilibrium at the
neutrinosphere, evaluated at the resonance momentum.
The peak in \(\sin 2 \theta_m\) at the resonance momentum
gives the derivative terms,
while \(\sin 2 {\bar \theta}_m\) yields the other terms.

In the adiabatic limit,
the effects of the neutrino background on the flavor diagonal elements
can be incorporated by using in \(\gamma\) a new,
effective electron density
\(N'_e ( r, \phi ) \equiv N_e ( r ) + N_{Ad} ( r, \phi) \), analogous
to the nonadiabatic case.
For typical supernovae conditions, \(N_{Ad}\)
is about 2.5 times larger than the nonadiabatic neutrino density,
\(N_{NA}\), and has the opposite sign from the electron density.
Thus the effective electron density will be considerably less
than the physical electron density---especially for large \(\phi\)
or late times or large r.
Similarly, the effects of the neutrino background on the flavor off-diagonal
elements can be incorporated by using in \(\gamma\) a new, effective
vacuum mixing.  For \(\sin^2 2 \theta << 1\)
the resonance is narrow so Eq. (\ref{Vem})
can be evaluated at the resonance position, \(r_0\), to yield
\begin{equation}
\sin 2 \theta'(\phi) \equiv \sin 2 \theta [ 1 -
({ N_{Ao}(r_0,\phi)  \over N'_e (r_0,\phi) }) ]
\label{s'}
\end{equation}
The dominant terms in \(N_{Ao}\) are those with the large logarithmic factor.
Evaluating at the average muon-neutrino energy,
and using the neutrino densities
given in Ref. \cite{PRL} yields that
\((N_{Ao}/N'_e)|_{r_0} > 1 \) for \(\Delta < 700\) eV\(^2\),
with a maximum of 8 at \(\Delta \sim 10\) eV\(^2\).
Thus the neutrino background is no longer `small',
for \(\Delta < 700\) eV\(^2\).

Of the two limits studied here, the adiabatic limit
is the most relevant for estimating the effects of a small neutrino
background on supernovae nucleosynthesis.
This is because when the neutrino background is neglected
the flavor evolution is extremely adiabatic for almost all
of the relevant parameter space,
and also because neutrino masses are only probed
if significant amounts of flavor conversion occur.
As \(\Delta\) approaches 700 eV\(^2\) from above, the 'small`
neutrino background increases \(P_c\), driving the system
away from the adiabatic limit.
This does not imply that the nonadiabatic limit is then reasonable
but instead that the oscillatory terms in Eq. ({\ref{ro}}) become important.
This is because the MSW effect dramatically enhances the
amplitude of the oscillatory terms over that of
the other flavor off-diagonal terms
so that only a very small amount of coherence is required for the oscillatory
terms to dominate.  Then the neutrino and antineutrino flavors evolve
nonlinearly and considerable further study is required
in order to reliably describe them.
For \(\Delta < 700\) eV\(^2\) the connection between
r-process nucleosynthesis and cosmologically relevant neutrino masses
is enervated.

I acknowledge
Y.Z. Qian, G. Fuller, B. Balantekin and W. Haxton
for useful discussions, and for pointing out a mistake in
an earlier version of this paper.
This work is supported in part by
the U.S. Department of Energy under Grant No. DE-FG02-91ER40661.
I also thank the Institute for Nuclear Theory at the University
of Washington for its hospitality and partial support.

\raggedbottom

\raggedbottom

\end{document}